\documentstyle[12pt,fleqn]{article}
\begin{document}

\vskip 0.15in

\title{Must Time Machine Be Unstable against Vacuum Fluctuations?}

\vskip 0.2in
\author{Li-Xin Li\thanks{Email address: lixin@eden.rutgers.edu} \\ CCAST (World Laboratory) P.O.Box 8730, Beijing
100080 and \\Department of Physics, Beijing Normal University,\\
Beijing 100875, China and\\ Institute of
Theoretical Physics, Academia Sinica, \\Beijing 100080, China}

\maketitle

\begin{abstract}

In the appearance of absorption material, the quantum vacuum
fluctuations of all kinds of fields may be smoothed out and the
spacetime with time machine may be stable against vacuum fluctuations.
The chronology protection conjecture might break down, and the anti-chronology
protection conjecture might hold: There is no law of physics preventing
the appearance of closed timelike curves.

~~~

PACS number(s): 04.20.Gz, 04.62.+v

\end{abstract}

\section*{}
Whether or not the laws of physics permit the existence of closed
timelike curves (time machines) is one of important problems in the
research field of modern general relativity [1-4]. Hawking has suggested
the chronology protection conjecture which states that the laws of
physics do not allow the appearance of closed timelike curves [5]. This
conjecture was based on the fact that several calculations about the
vacuum fluctuations in the spacetime containing closed timelike curves
reveal that the renormalized stress-energy tensor always diverges at the Cauchy
horizon which separates the region with closed timelike curves from the
region without closed causal curves, and also diverges at the polarized
hypersurfaces which are nested inside the Cauchy horizon [1,6-10].
Recently Li et al have challenged the chronology protection conjecture
[11,12]. They found that in some cases the renormalized stress-energy tensor of
vacuum fluctuations might be convergent at the Cauchy horizon and the
polarized hypersurfaces. In this paper I will show that the appearance
of absorption material in the spacetime containing closed timelike
curves can cause that the renormalized
stress-energy tensor of vacuum fluctuations should be
convergent at the Cauchy horizon and the polarized hypersurfaces, and
such a spacetime might be stable against vacuum fluctuations.

Research of high-energy scattering reveals that, if the energy
$\hbar\omega$ of the
incident particle is sufficiently large (the extreme relativistic
limit), for many processes the total cross section $\sigma_t$ increases
with $\hbar\omega$, or tends to a constant. Examples are: 1). The
electron-positron pair production by a photon in the photon-electron
collision
\begin{eqnarray}
\sigma_t=\alpha r_e^2({28\over9}\ln{2\hbar\omega\over m_ec^2}-10.8),
~~~~\hbar\omega\gg m_ec^2,
\end{eqnarray}
where $\hbar\omega$ is the energy of incident photon,
$m_e$ is the mass of the electron, $r_e=e^2/4\pi m_ec^2$ is the
classical radius of the electron, and $\alpha$ is the fine structure
constant [13]; 2). The cross section for electron energy loss through
bremsstrahlung
\begin{eqnarray}
\sigma_t=4Z^2\alpha^3({\hbar\over m_ec})^2(\ln{2\hbar\omega\over
m_ec^2}-{1\over3}),
~~~~\hbar\omega\gg m_ec^2,
\end{eqnarray}
for bare nuclei, where $\hbar\omega$ is the energy of the incident
electron, $Ze$ is the charge of the nuclei [14]. If the screening is
considered, the cross section is [14]
\begin{eqnarray}
\sigma_t=4Z^2\alpha^3({\hbar\over m_ec})^2[\ln(183Z^{-1/3})+{2\over9}],
~~~~\hbar\omega\gg m_ec^2\alpha^{-1}Z^{-1/3}.
\end{eqnarray}
3). For high energy inclusive neutrino-nucleon reaction
\begin{eqnarray}
\sigma_t\sim{G_F^2M\hbar^3\omega\over\pi c^4},
\end{eqnarray}
where $\hbar\omega$ is the laboratory energy of the initial neutrino,
$M$ is the mass of the target nucleon, $G_F$ is the Fermi beta-decay
constant [14,15]; 4). The Drell-Yan process of muon pair production in
proton-proton collision
\begin{eqnarray}
\sigma_t={1\over3}{4\pi\alpha^2e_q^2\over3}({\hbar\over M_\mu c})^2,~~~
{\rm as}~~\hbar\omega\rightarrow\infty,
\end{eqnarray}
where $e_q$ is the charge of the quark in units of the electric charge,
$M_\mu$ is the mass of muon pair, $\hbar\omega$ is the energy of
the incident proton [16]. This is a common feature of hadron collision
with high energy: the total cross section tends to a constant
\begin{eqnarray}
\sigma_t\sim({\hbar\over m_\pi c})^2,
\end{eqnarray}
where $m_\pi$ is the mass of the pion [14,17]; 5). The pair production
by graviton collisions
\begin{eqnarray}
\sigma_t={38\pi\over3}{G^2(\hbar\omega)^2\over c^8},~~~~\hbar\omega\gg
mc^2,
\end{eqnarray}
where $\hbar\omega$ is the energy of the graviton, $G$ is the Newton
gravity constant, $m$ is the mass of the particle pair [18]. The total cross
section for annihilation of a pair of pions into gravitons at high
energy is identical with (7) with $\hbar\omega$ as the energy of the
pair of particles [18]. The cross section for gravitational scattering
of identical particles is similar to (7)
\begin{eqnarray}
\sigma_t\sim{G^2(\hbar\omega)^2\over c^8},~~~~\hbar\omega\gg
mc^2,
\end{eqnarray}
here $\hbar\omega$ is the energy of the incident particle and $m$ is its
mass [18]. By theses facts, it seems that materials will become opaque
at extremely high energy of incident particles or waves (including
gravitons or gravitational fields), since in such a case the absorption
caused by various types of scattering processes becomes very important.
Especially, Eq.(8) might act as {\em an universal cross section for
absorption} caused by scattering of all kinds of incident waves,
including the scalar, the spinor, the vector, and the tensor waves
(including the gravitational
waves), since all particles are subject to the gravitation. For the
discussion in this paper it is convenient to write Eqs.(1-8) into three
classes:

i). $\sigma_t$ increases with $\omega$ through
\begin{eqnarray}
\sigma_t=\alpha\omega^s,~~~~s>0~~{\rm and}~~\omega\gg\omega_0,
\end{eqnarray}

ii). $\sigma_t$ increases with $\omega$ through
\begin{eqnarray}
\sigma_t=s\ln{\omega\over\omega_0},~~~~s>0~~{\rm
and}~~\omega\gg\omega_0,
\end{eqnarray}
$\omega_0$ in Eqs.(9) and (10) are some critical values of $\omega$.

iii). $\sigma_t$ tends to a constant as $\omega\rightarrow\infty$
\begin{eqnarray}
\sigma_t\rightarrow\pi r_0^2.
\end{eqnarray}

In the spacetime with closed timelike curves, the regularized Hadamard
function for the scalar field is given by the expansion in the
geometric-optics approximation [1,19,20]
\begin{eqnarray}
G^{(1)}_{\rm
reg}(x,x^\prime)=\sum_{N=1}^\infty{\Delta_N^{1/2}\over2\pi^2}({1\over\sigma_N}+
v_N\ln\vert\sigma_N\vert+w_N),
\end{eqnarray}
where $\sigma_N$, the $N$-th geodetic interval between $x$ and
$x^\prime$, is equal to $1/2$ the square of the proper distance along the
geodesic ${\cal G}_N$ from $x^\prime$ to $x$ with the winding number $N$,
multiplied by $-1$ if ${\cal G}_N$ is timelike and $+1$ if it is
spacelike; $\Delta_N$, $v_N$, and $w_N$ are regular functions of $x$ and
$x^\prime$. For the spinor field, the vector field and the tensor field
(including the gravitational field), the Hadamard functions are similar
to (12) [8,18,22]. We are only interested in the $1/\sigma_N$ terms in
(12), since they are most important for the divergence of the
renormalized stress-energy tensor of the vacuum fluctuations in the
case without absorption material. Therefore we only deal with
\begin{eqnarray}
G^{(1)}_{\rm
reg}(x,x^\prime)\sim\sum_{N=1}^\infty{\Delta_N^{1/2}\over2
\pi^2}{1\over\sigma_N},
\end{eqnarray}
for all kinds of quantum fields. With the method like that used by Kim
and Thorne in [1], in the appearance of absorption material,
$\Delta_N^{1/2}$ can be evaluated through
\begin{eqnarray}
\Delta_N^{1/2}\simeq\Delta_N^{(0)1/2}\Gamma^N,
\end{eqnarray}
where $\Gamma$ is the transmission coefficient through the material and
$\Delta_N^{(0)}$ is the $\Delta_N$ function without the appearance of
absorption material. If the absorption material is so dense that the
cross section overlaps each other, we have $\Gamma=0$,
otherwise $\Gamma$ is given by
\begin{eqnarray}
\Gamma=e^{-nz\sigma_t},
\end{eqnarray}
where $\sigma_t$ is the total cross section, $n$ is the number density
of target particles inside the material, and $z$ is the thickness of the
material [21]. The renormalized stress-energy tensor can be computed
from $G_{\rm reg}^{(1)}$ by acting it with some second-order differential
operator [8,23,24]
\begin{eqnarray}
\langle T_{\mu\nu}\rangle_{\rm ren}\sim-\sum_{N=1}^\infty{1\over6\pi^2}
{\Delta_N^{(0)1/2}\Gamma^N\over\sigma_N^3}t_{\mu\nu,N},
\end{eqnarray}
where $t_{\mu\nu,N}$ is a regular tensor [4], $\sigma_N=\sigma_N(x,x)$,
$\Delta_N^{(0)}=\Delta_N^{(0)}(x,x)$. The metric perturbations created
by vacuum fluctuations can be estimated through [1,4]
\begin{eqnarray}
\delta g_{\mu\nu}\sim\sum_{N=1}^\infty\Delta_N^{1/2}{l_{\rm
P}^2\over\sigma_N}\sim\sum_{N=1}^\infty\Delta_N^{(0)1/2}{l_{\rm
P}^2}{\Gamma^N\over\sigma_N},
\end{eqnarray}
where $l_{\rm P}$ is the Planck length.

Near the Cauchy horizon and the polarized hypersurfaces, the average
frequency of the wave package propagating in the spacetime has the
magnitude
\begin{eqnarray}
\overline\omega\sim{cL\over\sigma_1},
\end{eqnarray}
where $L$ is the spatial distance between the identified points in the
frame rest relative to the absorption  material,
$\overline\omega$ is very large since near the horizon or the polarized
hypersurfaces we have $\sigma_N\sim0$. Now
I discuss the three cases listed above respectively:

{\sl Case i)}. Inserting (18) into (9), we have
\begin{eqnarray}
\sigma_t\sim\alpha (cL)^s\sigma_1^{-s},
\end{eqnarray}
From (15), (16) and (19) we have
\begin{eqnarray}
\langle T_{\mu\nu}\rangle_{\rm ren}\sim-\sum_{N=1}^\infty{1\over6\pi^2}
\Delta_N^{(0)1/2}\sigma_1^{-3}\exp[-N\alpha
nz(cL)^s\sigma_1^{-s}]t_{\mu\nu,N},
\end{eqnarray}
where we have used $\sigma_N\sim\sigma_1$ [1,4]. For $s>0$, we have
$\langle T_{\mu\nu}\rangle_{\rm ren}\rightarrow0$ as
$\sigma_1\rightarrow0$ (approaching the Cauchy horizon or the polarized
hypersurfaces). Therefore the renormalized stress-energy tensor of the vacuum
fluctuations does not diverge at the Cauchy horizon and the polarized
hypersurfaces. From (15), (17) and (19) we have
\begin{eqnarray}
\delta g_{\mu\nu}\sim\sum_{N=1}^\infty\Delta_N^{(0)1/2}{l_{\rm
P}^2\over\sigma_1}\exp[-N\alpha nz(cL)^s\sigma_1^{-s}],
\end{eqnarray}
As $\sigma_1\rightarrow0$, we have $\delta g_{\mu\nu}\rightarrow0$, the
maximum value of $\delta g_{\mu\nu}$ is at $\overline\sigma_1\sim
cL(\alpha nz)^{1/s}$ and
\begin{eqnarray}
(\delta g_{\mu\nu})_{\rm max}\sim\Delta_1^{(0)1/2}{l_{\rm
P}^2\over\overline\sigma_1}\sim \Delta_1^{(0)1/2}l_{\rm
P}^2(cL)^{-1}(\alpha nz)^{-1/s}.
\end{eqnarray}
If this value is much smaller than $1$, the spacetime will be stable
against vacuum fluctuations. If we take $\sigma_t$ given by Eq.(4), then
$s=1$ and $\alpha=G_F^2M\hbar^3/\pi c^4$, and
\begin{eqnarray}
(\delta g_{\mu\nu})_{\rm
max}\sim\Delta_1^{(0)1/2}G_F^{-2}M^{-1}\hbar^{-3}c^{3}l_{\rm
P}^2(nzL)^{-1}.
\end{eqnarray}
If $\Delta_1^{(0)}\sim1$, $M\sim m_e$, then $(\delta g_{\mu\nu})_{\rm
max}\sim10^{-44}(nzLl_{\rm P})^{-1}\ll1$ for $nzL\gg10^{-11}{\rm
cm}^{-1}$. This can be satisfied for a large range of $z, L$ and $n$.
For example, we take $L\sim z\sim10^2{\rm
cm}$, $n\sim10^4$cm$^{-3}$, then $(\delta g_{\mu\nu})_{\rm
max}\sim10^{-19}$. If $\sigma_t$ is given by (8) (this is an important
case since we expect that this is an universal absorption cross section),
then $s=2$ and $\alpha\sim G^2\hbar^2c^{-8}\sim l_{\rm P}^4c^{-2}$, and
\begin{eqnarray}
(\delta g_{\mu\nu})_{\rm
max}\sim\Delta_1^{(0)1/2}(nzL^2)^{-1/2},
\end{eqnarray}
which is much smaller than $1$ if $nzL^2\gg\Delta_1^{(0)}$, this can
also be satisfied for a large range of $z, L$ and $n$. For example,
if $\Delta_1^{(0)}\sim1$, $L\sim z\sim10^5{\rm cm}$,
$n\sim10^5{\rm cm}^{-3}$, then $
(\delta g_{\mu\nu})_{\rm max}\sim10^{-10}$, this is a much smaller value
which cannot destroy the background spacetime. For $n\sim 10^5{\rm
cm}^{-3}$, the mass density is $\rho\sim10^{-19}{\rm g}/{\rm cm}^3$. The
influence of the material with such a small density on the background
spacetime can also be neglected.

{\sl Case ii)}. Inserting (18) into (10), we have
\begin{eqnarray}
\sigma_t\sim s\ln(cL\omega_0^{-1}\sigma_1^{-1}).
\end{eqnarray}
From (15), (16) and (25) we have
\begin{eqnarray}
\langle T_{\mu\nu}\rangle_{\rm ren}\sim
-\sum_{N=1}^\infty{1\over6\pi^2}
\Delta_N^{(0)1/2}(cL\omega_0^{-1})^{-nzs}\sigma_1^{nzs-3}
t_{\mu\nu,N},
\end{eqnarray}
and we find $\langle T_{\mu\nu}\rangle_{\rm ren}\rightarrow0$ as
$\sigma_1\rightarrow0$ if $nzs>3$. From (15), (17) and (25) we find
\begin{eqnarray}
\delta g_{\mu\nu}\sim\sum_{N=1}^\infty\Delta_N^{(0)1/2}l_{\rm
P}^2(cL\omega_0^{-1})^{-nzs}\sigma_1^{nzs-1},
\end{eqnarray}
and $\delta g_{\mu\nu}\rightarrow0$ as $\sigma_1\rightarrow0$ if
$nzs>1$. Therefore, if
\begin{eqnarray}
nzs>3,
\end{eqnarray}
both $\langle T_{\mu\nu}\rangle _{\rm ren}$
and $\delta g_{\mu\nu}$ tends to
zero as $\sigma_1\rightarrow 0$. Since (10) holds for
$\omega\gg\omega_0$ or $cL\omega_0^{-1}\sigma_1^{-1}\gg1$, we can
estimate the maximum value of $\delta g_{\mu\nu}$ by taking
$cL\omega_0^{-1}\sigma_1^{-1}=1$ or
$\sigma_1=\overline\sigma_1\equiv cL\omega_0^{-1}$, then
\begin{eqnarray}
(\delta g_{\mu\nu})_{\rm max}\sim\Delta_1^{(0)1/2}l_{\rm
P}^2\omega_0(cL)^{-1}.
\end{eqnarray}
For $\sigma_t$ given by Eq.(1), we have $s\sim\alpha r_e^2$ and
$\omega_0\sim m_ec^2/\hbar$, then
\begin{eqnarray}
(\delta g_{\mu\nu})_{\rm max}\sim\Delta_1^{(0)1/2}l_{\rm
P}^2(r_eL)^{-1},
\end{eqnarray}
and $nzs\sim\alpha r_e^2nz$, Eq.(28) demands
\begin{eqnarray}
\alpha r_e^2nz>3.
\end{eqnarray}
If $\Delta_1^{(0)}\sim1$ and (31) is satisfied, we always have
$(\delta g_{\mu\nu})_{\rm
max}\ll1$ for any $L$ with macroscopic scale. Eq.(31) can be satisfied for a
large range of $z$ and $n$. For example, we take $z\sim10^4{\rm cm}$,
then (31) holds for $n>10^{24}{\rm cm}^{-3}$ or
$\rho>1{\rm g}/{\rm cm}^{-3}$. This can be
satisfied for laboratory materials. For the cross section given by
Eq.(2), we have similar results. These mean that the vacuum
fluctuations of electromagnetic field and electron (positron) field can
be smoothed out by introducing absorption material with mass density
greater than $1$g/cm$^3$.

{\sl Case iii)}. For the cross section given by (11) it seems impossible
to smooth out the vacuum fluctuations by this kind of scattering
since from (15) we obtain
$\Gamma=$constant. However if the absorption material is so dense that
the cross section of each target particle overlaps each other, or $d\sim
r_0$ where $d$ is the average distance between the neighboring target
particles, then (15) breaks down and we have $\Gamma=0$, and both $\langle
T_{\mu\nu}\rangle _{\rm ren}$ and $\delta g_{\mu\nu}$ vanishes in such a case.
For the cross section given by (6) we have $d\sim r_0\sim\hbar/m_\pi
c\sim 10^{-13}{\rm cm}$, such materials have the same mass density as
the neutron stars. For such a high density one may argue that the
gravitational field produced by the absorption material will be very
great so that the great tidal force will prevent people from time
travel. However, if the Schwarzschild radius $r_g$ of the star of the
absorption material is very small comparing with the radius $R$ of
the star, the gravitational field out side the star produced by itself
will be very small and the time traveler can travel in time bypassing
the star along a nongeodetic closed timelike curve [12]. $r_g\ll R$ is
satisfied if $R\ll c/(G\rho)^{1/2}\sim10^{7}{\rm cm}$ if
$\rho\sim10^{15}{\rm g}/{\rm cm^3}$. To prevent the star from collapsing
into a black hole, $R$ should satisfy $R<(M_{\rm
max}/\rho)^{1/3}\sim10^6{\rm
cm}$ where $M_{\rm max}$ is the maximum mass of the stable neutron
stars. If we take $R\sim10^4{\rm cm}$, the gravitational field outside
the star produced by itself should be very small since $r_g/R\sim10^{-6}$
and such a star should be stable. (If Eq.(8) does be a universal cross
section for absorption, material with such a high density will not be
necessary for smoothing out the vacuum fluctuations.)

From above discussions, we find that if absorption material with
appropriate density is introduced, all vacuum fluctuations may be
smoothed out such that the metric perturbation created by vacuum
fluctuations will be very small and the spacetime with time machine may
be stable
against vacuum fluctuations. Eq.(8) may be an universal cross section for
absorption of all kinds of fields since we know that all particles are
subject to the gravitation. If this is correct, then the
quantum vacuum fluctuations of all kinds of fields may be smoothed out
in the appearance of absorption material with the number density greater
than $n_0=(zL^2)^{-1}$, or the mass density greater than
$\rho_0=m_n/zL^2$ where $m_n$ is the mass of nuclei. If we take $z\sim
L\sim10^4{\rm cm}$, then $\rho_0\sim10^{-36}{\rm g}/{\rm cm}^3$,
the average mass density of the cosmic galaxies is greater than this
value.

The conclusion is: The renormalized stress-energy tensor of vacuum
fluctuations may be smoothed out by introducing absorption material, such
that the spacetime containing time machines may be stable against vacuum
fluctuations of all kinds of fields (including the gravitational field).
Contrary to the chronology protection conjecture, I give the {\em
anti-chronology protection conjecture: There is no law of physics
preventing the appearance of closed timelike curves}.

I am very thankful to L. Liu for helpful discussions.

\end{document}